%% file: main.tex
\documentclass[a4paper,12pt,onecolumn]{article}

\usepackage[utf8]{inputenc}
\usepackage[T1]{fontenc}

\usepackage[english]{babel}

\usepackage{FAS}
\usepackage{graphicx} 
\usepackage{csquotes} 
\usepackage{xcolor} 
\usepackage{subfig} 
\usepackage{ragged2e}
\usepackage{amsthm}
\usepackage{verbatim}
\usepackage{array}
\usepackage{caption}
\usepackage{hyperref}
\usepackage{enumerate}
\usepackage[shortlabels]{enumitem}
\usepackage{microtype}
\usepackage[all]{nowidow}
\usepackage{tikz}
\usepackage{tubscolors}
\usepackage{cleveref}
\usepackage{tikz-uml}
\usetikzlibrary{%
	shapes,
	automata,
	arrows,
	matrix,
	backgrounds,
	fit,
	patterns,
	decorations.markings, 
	svg.path, 
	shapes.multipart, 
	shadows, 
	fadings,
	shadings,
	positioning, 
	calc, 
	decorations, 
	decorations.text,
	intersections, 
	angles, 
	quotes,
	decorations.pathmorphing, 
	arrows.meta, 
	patterns, 
	decorations.pathreplacing, 
	calligraphy,
	snakes,
	bending,   			 	 
	babel,
        external,
	spath3%
}     		 	

\tikzexternaldisable 

\usepackage{tikz-3dplot}
\usepackage{tkz-euclide}
\usepackage{pgfplots}
\usepackage{pgfplotstable}
\usepgfplotslibrary{fillbetween}
\usetikzlibrary{ext.paths.ortho}

\input{shapes}
\input{tikzstyles}

\usepackage[
			bibstyle=ieee,
			citestyle=numeric-comp,
			backend=biber,
			minnames=1,
			maxcitenames=2,
			maxbibnames=10,
			isbn=false,
			eprint=false,
			url=true,
			natbib=true,
			clearlang=true,
			date=year
			]{biblatex} 

\DeclareFieldFormat[unpublished]{annotation}{}
\DeclareFieldFormat[misc]{url}{\mkbibacro{URL}\addcolon\space\url{#1}} 
\DeclareFieldFormat[misc]{url}{\mkbibacro{report}\addcolon\space\url{#1}} 
\DeclareListFormat[inproceedings]{eventtitle}{} %
\DeclareListFormat{language}{}           

\DeclareDatamodelEntrytypes{standard}
\DeclareDatamodelEntryfields[standard]{type,number}
\DeclareBibliographyDriver{standard}{%
  \usebibmacro{bibindex}%
  \usebibmacro{begentry}%
    \printfield{shorttitle}%
    \usebibmacro{prenote}%
    \printtext{\addcolon}%
    \printfield{year}%
  \setunit{\labelnamepunct}\newblock
  \usebibmacro{title}%
  \newunit\newblock
  \printnames{author}%
  \setunit{\addspace}\newblock
  \printfield[parens]{type}%
  \newunit\newblock
  \iftoggle{bbx:url}
    {\usebibmacro{url+urldate}}
    {}%
  \newunit\newblock
  \usebibmacro{addendum+pubstate}%
  \setunit{\bibpagerefpunct}\newblock
  \usebibmacro{pageref}%
  \newunit\newblock
  \usebibmacro{related}%
  \usebibmacro{finentry}}

 \DeclareCiteCommand{\citenormp}[\mkbibparens]%
{}%
{%
	\printfield{shorttitle}%
	\usebibmacro{prenote}%
	\printtext{\addcolon}%
	\bibhyperref{\printfield{year}}%
}%
{\multicitedelim}%
{%
	\usebibmacro{postnote}%
}%

    \DeclareCiteCommand{\citenormt}%
{}%
{%
	\printfield{shorttitle}%
	\usebibmacro{prenote}%
	\printtext{\addcolon}%
	\bibhyperref{\printfield{year}}%
}%
{\multicitedelim}%
{%
	\iffieldundef{postnote}{}%
	{%
		\addspace%
		\parentext{\printfield{postnote}}
	}%
}%

\DeclareCiteCommand{\citenormy}
{}
{%
	\printfield{shorttitle}%
	\printtext{\addcolon}%
	\bibhyperref{\printfield{year}}}%
{\multicitedelim}
{}

\setlist[enumerate]{nosep, itemsep=0.5em, topsep=0.75em}
\setlist[itemize]{nosep, itemsep=0.5em, topsep=0.75em}

\DeclareFieldFormat{urldate}{\bibstring{urlseen}\space#1}
\DeclareFieldFormat{url}{\url{#1}}

\DeclareBibliographyCategory{needsurl}
\newcommand{\entryneedsurl}[1]{\addtocategory{needsurl}{#1}}

\renewbibmacro*{url+urldate}{%
	\ifcategory{needsurl}
	{
		\usebibmacro{urldate}%
		\newunit
		\usebibmacro{url}%
	}
	{
		\ifentrytype{online}
		{
			\usebibmacro{urldate}%
			\newunit
			\usebibmacro{url}%
		}
		{
			\ifentrytype{thesis}
			{
				\usebibmacro{url}%
			}
			{}       
		}    
	}
}

\entryneedsurl{siemensplmsoftware2018}

\usepackage{longtable}

\pgfdeclarelayer{background}%
\pgfdeclarelayer{pre main}%
\pgfdeclarelayer{foreground}%
\pgfsetlayers{background,connections,pre main,main,foreground}%

\begin{document}
%
\title{%
Towards Closing the Gap between Model-Based Systems Engineering and Automated Vehicle Assurance: Tailoring Generic Methods by Integrating Domain Knowledge
}
%
%
\author{Marcus Nolte%
\footnote{Corresponding author: m.nolte@tu-braunschweig.de}
\footnote{Institut für Regelungstechnik, Technische Universität Braunschweig, Braunschweig}
\ and Markus Maurer%
\footnotemark[2]
}
%
%
\date{}

\maketitle \thispagestyle{empty}

\begin{abstract}
    Designing, assuring and releasing safe automated vehicles is a highly interdisciplinary process.
    As complex systems, automated driving systems will inevitably be subject to emergent properties, i.\,e. the properties of the overall system will be more than just a sum of the properties of its integrated elements.
    Safety is one example of such emergent properties.
    In this regard, it must be ensured that effects of emergence do not render an overall system that is composed of safety-approved sub systems unsafe.

   The key challenges in this regard are twofold:
    Regarding the interdisciplinary character of the development and assurance processes, all relevant stakeholders must speak a common language and have a common understanding of the key concepts that influence system safety.
    Additionally, the individual properties of system elements should remain traceable to the system level.

    Model-Based Systems Engineering (MBSE) provides an interdisciplinary mindset, as well as methods and processes to manage emergent system properties over the entire system lifecycle.
    By this, MBSE provides tools that can assist the assurance process for automated vehicles.
    However, concepts from the domain of MBSE have a reputation for not being directly accessible for domain experts who are no experts in the field of Systems Engineering.

    This paper highlights challenges when applying MBSE methods to the design and development of automated driving systems. 
    It will present an approach to create and apply domain-specific SysML profiles, which can be a first step for enhancing communication between different stakeholders in the development and safety assurance processes.
\end{abstract}

\vspace{-0.75em}
\section[]{Motivation}
\label{sec:02_implemented-views}
\vspace{-0.5em}
The challenges related to the assurance of automated driving systems, i.e. proving that automated vehicles do not pose an unreasonable risk to its passengers and other traffic participants, are still manifold.
Research questions remain all along the process, e.g., from defining risk acceptance criteria over performing effective safety analyses to the effective verification and validation of AI-based system components.

A challenge in this context is that different stakeholders in the assurance process come from different disciplines with different qualifications and different mindsets. 
Safety- and System Engineers must, e.g., work together with AI-experts, function developers and/or Test Engineers.
As such, \textcite{koopman2017} highlight that safety engineering must be treated as a highly interdisciplinary process that must solve a ``set of coupled problems [...] in a coordinated, cross-domain manner.'' \parencite[91]{koopman2017} 

This is a challenge that Systems Engineering is supposed to solve: The International Council on Systems Engineering (INCOSE) defines Systems Engineering as ``a transdisciplinary and integrative approach to enable the successful realization, use, and retirement of engineered systems [...]'' \parencite{walden2023}.
A central concern in Systems Engineering processes is the management of complexity and emergent system properties. 
For this purpose, \emph{Model-Based} Systems Engineering (MBSE)\footnote{MBSE combines Systems Engineering processes with (semi-) formal models (e.g., defined in SysML \parencite{omgsysml2019}) for modeling requirements, architectures, test specifications, and other process artefacts.} provides (semi-) formal methods for modeling the system and related elements.
These models facilitate traceability between the different artefacts which are generated during design, development, and assurance.

While MBSE-based methods and processes show great potential for rigorous development and assurance processes in the domain of automated driving, there are significant adoption challenges in industry settings~\parencite{blott2023}.
Besides typical ressentiments such as the steep learning curve and the modeling overhead that come with MBSE-adoption processes, \textcite{blott2023} discuss the importance of introducing domain knowledge into the modeling process.
Furthermore, model-based methods cannot ensure the \emph{completeness} of the generated models.
Also, scalability can be questioned, when the models become larger and more complex.
However, model-based approaches can help significantly to ensure the consistency of the modeled artefacts.
In this regard, integrating domain knowledge can help to make models more accessible for domain experts. 
This can facilitate communication between Systems Engineers and domain experts and by this indirectly help to capture relevant system properties and requirements.
Furthermore, applying domain knowledge can help to facilitate the integration of domain-specific tooling with the models.

We will outline an approach to tailor established (MB)SE processes using domain knowledge in the field of automated driving by creating domain-specific SysML profiles.
The approach is based on a comparative analysis of selected standards in the field of automated driving and the central process standard for Systems- and Software Engineering (\citenormt{iso15288-23}).
Based on this analysis, examples for common concepts from those standards in both domains will be extracted and documented in a corresponding concept model.
This concept model will be used as a basis for creating an example of a domain-specific SysML profile which can be used to model AD-specific artefacts (or work products) that can contribute to a traceable system specification.
\vspace{-0.75em}
\section{Related Work}
\label{sec:related-work}
\vspace{-0.5em}
Recent research \parencite{meyer2022, kinay2024, sahin2021, raulf2023} and industry position papers \parencite{siemensplmsoftware2018} show that the application of MBSE for the design and assurance of automated vehicles is gaining traction.

\textcite{meyer2022} describe an MBSE-driven approach for deriving a logical architecture from use case and scenario descriptions.
They apply MBSE processes to domain-specific ideas of scenario-based development for automated vehicles, creating models that reflect a model-based and SOTIF-compliant assurance case in Goal Structuring Notation.
\textcite{kinay2024} focus on an MBSE-supported generation of scenarios for verification and validation.
\textcite{raulf2023} as well as \textcite{sahin2021} apply an MBSE-based architecture framework to the definition of vehicle concepts and mobility services offered by automated vehicles.

The authors in all examples apply principles from MBSE without domain-specific tailoring of the required concepts, as suggested by \textcite{blott2023}. 
The authors of the Systems Architecture Framework (SAF \parencite{leute2021})\footnote{See modeling approach at \url{https://github.com/GfSE/SAF-Specification/blob/main/developing-saf/development.md} (visited 09/26/24).} show how a traceable definition of concept models and SysML profiles for domain-specific modeling can be achieved.

\section{Translating from Systems Engineering Concepts to Domain-Specific Concepts}
\label{sec:se-to-de}
\vspace{-0.5em}
A plethora of different concepts has been introduced in the field of automated driving to address the needs of the various stakeholders in the development and assurance process.
The \emph{Operational Design Domain} (\citenormt{iso34503}), behavioral competencies \parencite{avsc2021} or a particular domain-specific understanding of \emph{Scenarios} (\citenormt{iso21448}, \citenormt{iso34501}) are prominent examples of such concepts.
Regarding safety assurance, \citenormt{iso2018} and \citenormt{iso21448} (SOTIF) have introduced  domain-specific artefacts and process steps that are mandatory to create and perform to claim standard compliance.

In the field of Systems Engineering, \citenormt{iso15288-23} defines a generic Systems Engineering process framework for managing the entire system life cycle.
The standard rigorously defines processes in an input-process-output (IPO) scheme, meaning that not only the process steps themselves are defined, but also the required input artefacts to perform the processes and their expected output artefacts.

\begin{figure}[t]%
\centering
\input{fusa-sotif-mapping.tikz}
\caption[Mapping]{Mapping of safety-related artefacts from \citenormt{iso21448} and \citenormt{iso2018} to \citenormt{iso15288-23} processes (matching colors indicate that an artefact is generated in the corresponding process).\footnotemark{}}%
\label{fig:mapping_highlevel}
\end{figure}
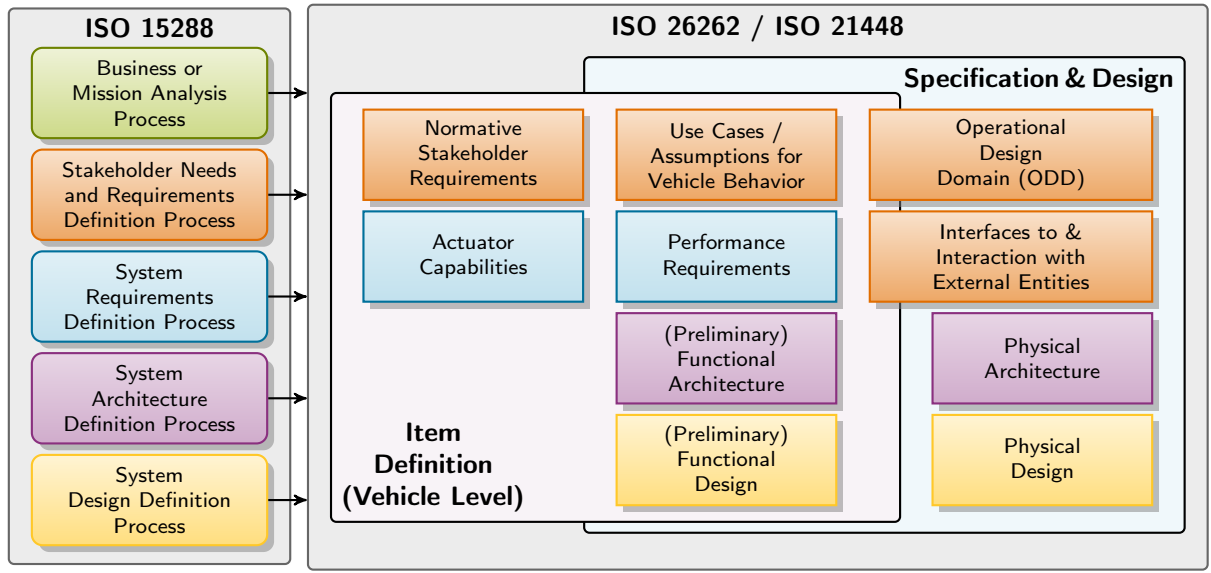

The relevant automotive standards do, in general, not refer to details of the generic Systems Engineering standards such as \citenormt{iso15288-23}.
However, when comparing the processes and artefacts defined by \citenormt{iso15288-23} with the different terms and concepts in the domain-specific standards such as \citenormt{iso34503}, \citenormt{iso2018} or \citenormt{iso21448}, it becomes apparent that while the automotive domain has introduced its own terminology, there is a strong overlap between the underlying concepts. \Cref{fig:mapping_highlevel} shows how artefacts that are required by \citenormt{iso2018} and \citenormt{iso21448} can be mapped to processes defined by \citenormt{iso15288-23}.
\footnotetext{The Item Definition is considered at the vehicle level, the domain-specific wording is adapted to summarize concepts of \citenormt{iso2018}, \citenormt{iso21448}. Note that the ISO~3450X standards family gives additional (domain-specific) terminology proposals for some of the mentioned artefacts, e.g., regarding scenarios (\citenormt{iso34501}) or the ODD (\citenormt{iso34503}).}

To design and maintain an actual product from an initial business goal of an organization, \citenormt{iso15288-23} defines processes and support processes, among others, a) to capture the organization's business mission (\emph{Business or Mission Analysis Process}), b) to align this business mission with customer (or more general ``stakeholder'') needs (\emph{Stakeholder Needs and Requirements Definition Process}), c) to derive concrete requirements from those stakeholder needs and requirements (\emph{System Requirements Definition Process}), d) to define the (functional, logical, and physical) system architecture (\emph{System Architecture Definition Process}), and e) to assign (or allocate) requirements to architectural elements into a coherent system design (\emph{System Design Definition Process}).\footnote{Note that the \citenormt{iso15288-23} defines additional processes that target the entire system life cycle. For the sake of the argument, we will focus on the processes that can be aligned with the \emph{Design Phase} according to \citenormt[][-3]{iso2018}.} 

In the automotive domain, the central document that initiates the \citenormy{iso2018} reference process is the \emph{Item Definition}.
The analogous document within the \citenormy{iso21448} is the \emph{Specification and Design}.
Comparable artefacts between both documents are displayed in \Cref{fig:mapping_highlevel}:
According to \citenormt[-3][pp.~4f.]{iso2018}, among others, the item definition should be based on:
\begin{enumerate}[itemsep=0.2em, topsep=0.4em]
	\item legal requirements, national \& international standards (which can be summarized as normative stakeholder requirements),
	\item (assumed) actuator potentials (e.g., maximum torque, forces, etc.),\label{it:26262-1}
	\item assumptions regarding the item's behavior at vehicle level and the ``operational scenarios'' which concern the item's functionality,\label{it:26262-2}
	\item requirements (quality, performance, availability) regarding the item's functionality,\label{it:26262-3}
	\item the item's boundary, its interfaces and assumptions regarding its interaction with other items and elements (which can be summarized as a preliminary functional or logical architecture),\label{it:26262-4}
	\item the allocation and distribution of functions among the involved systems and elements (combined with existing functional requirements \citenormp[-3][Clause~5.4.1, p.~4, NOTE~2]{iso2018}; this corresponds to a preliminary functional design)\label{it:26262-5}.
\end{enumerate}

In different formulations, the points \ref{it:26262-1} to \ref{it:26262-5} can also be found inside the specification and design according to \citenormy{iso21448}.
The description of assumptions regarding the item's behavior at the vehicles and the operational scenarios are summarized as use cases, development scenarios, and the interaction of the automated driving system with external entities.
For SOTIF, this also includes the definition of an initial (rudimentary) Operational Design Domain.
For the specification and design, \citenormy{iso21448} also expects performance targets for ``sensors, controllers and actuators or other [System-] inputs and components'' \citenormp[Clause~5.2, p.~21]{iso21448} which includes items \ref{it:26262-1} and \ref{it:26262-2} above.
Finally, SOTIF demands a definition of ``system and vehicle architectures implementing the intended functionality'' and of ``the design of the relevant system and its elements implementing the intended functionality'' \citenormp[Clause~5.2, pp.~21f.]{iso21448}.
These requirements regarding architectures and design include the bullets \ref{it:26262-4} and \ref{it:26262-5}, but require additional \emph{technical} architectures and designs compared to the \emph{functional} architecture and design required for the \citenormy{iso2018} item definition.

\Cref{fig:mapping_highlevel}, which maps the artefacts described above to the \citenormy{iso15288-23}-processes, shows that stakeholder needs and requirements do not play central roles in \citenormy{iso2018} or \citenormy{iso21448}:
\citenormy{iso2018} focuses on normative stakeholder \emph{requirements} when are derived from regulation and standards, SOTIF does not explicitly mention stakeholder needs or requirements at all.
At the same time, SOTIF demands the definition of (risk) acceptance criteria, which from a Systems Engineering perspective, cannot be defined without proper stakeholder needs analyses.

Regarding the definition of use cases, assumptions about the vehicle behavior, the definition of an Operational Design (or Targeted Operational) Domain (ODD / TOD), as well as the definition of interfaces of the automated driving system (or vehicle-level Item for \citenormy{iso2018}) constitute parts of an Operational Concept\footnote{The Operational Concept is a document that aims at describing the purpose of a system, primarily from the perspective of assumed stakeholders. \citenormp{iso15288-23}} (OpsCon) according to \citenormy{iso15288-23}.

The demanded preliminary requirements, the (preliminary or legacy) functional and physical architectures, as well as the (preliminary) functional and physical designs in \citenormy{iso2018} and \citenormy{iso21448} can be attributed to the \emph{System Requirements Definition}, the \emph{System Architecture Definition} and the \emph{System Design Definition Processes} in \citenormy{iso15288-23}.
All \citenormy{iso2018} and \citenormy{iso21448} processes related to hazard and risk assessments (HARA) can be attributed to \citenormy{iso15288-23} risk management processes (not displayed in \Cref{fig:mapping_highlevel}).

Regarding traceability between the different artefacts, \citenormt[-5][Clause~7.4.1.5, p.~10]{iso2018} and \citenormt[-6][Clause~7.4.2 a), p.~11]{iso2018} demand the traceability of hardware and software safety requirements to the hardware and software architectural designs, respectively.
\citenormy{iso21448} considers traceability as ``desirable'', but does not make specific requirements.
In contrast to that, \citenormy{iso15288-23} defines ``traceability documentation'' as a required artefact for each defined process.

\subsection*{Critical Assessment of Automotive Standards}
Summarizing, the comparison of \citenormy{iso15288-23} with the established automotive safety standards \citenormy{iso2018} and \citenormy{iso21448} shows that the generic Systems Engineering standard contains more demands regarding traceable system (architecture) design.
As some of the artefacts required by \citenormy{iso2018} and \citenormy{iso21448} can be unambiguously assigned to \citenormy{iso15288-23} process requirements, it seems promising to structure these common artefacts in a way such that they become compatible to established Systems Engineering practices and such that redundant work can be avoided.

At the same time, compared to established Systems Engineering practices, particularly early design phase artefacts that are connected to capturing and tracing stakeholder values, needs, and requirements are rarely  considered in the automotive safety standards.
This is explainable, as for established automotive systems, the main stakeholder has long been the customer, and automotive OEMs have historically been well-equipped to identify customer needs.
Even for simpler SAE-Level-2 systems, as the human driver always remains in control, it may not be immediately necessary to consider the more general Systems Engineering processes that are related to defining thorough operational concepts.
However, at the least when SAE-Level-4 systems are concerned, these systems impact a variety of stakeholder groups, most importantly the general public.

In this sense, closing this gap between a wide body of knowledge for the design of safe automotive systems and the more generic Systems Engineering processes have the potential to enable a) the required interdisciplinary collaboration described in \Cref{sec:02_implemented-views}, and b) establish frameworks for better communication with non-technical stakeholders.

\section{From Translated Concept Models to Domain-Specific Models}
\label{sec:basic-process}
Identifying common artefacts and allocating these artefacts to the generic Systems Engineering processes defined by \citenormy{iso15288-23} is only a first step for closing this aforementioned gap.
A key challenge remains in translating and structuring (or formalizing) the translated concepts between both fields.

As argued in \Cref{sec:02_implemented-views}, applying approaches from Model-Based Systems Engineering to the domain of automated driving is promising, when it comes to establishing traceability along the entire system life cycle.
For the implementation of MBSE-supported processes in the automated driving domain, there are two challenges that are closely related to concerns regarding the steep learning curve and the overhead that is often attributed \parencite{blott2023} to MBSE.
a) Oftentimes, experts who possess rich domain knowledge \emph{and} the required (meta-) modeling expertise are difficult to come by.
b) The modeling language is typically not tailored towards the specific application domains, causing communication challenges between System Engineers and domain experts.

\begin{figure}[htbp!]%
	\centering
	\includegraphics[width=.8\textwidth]{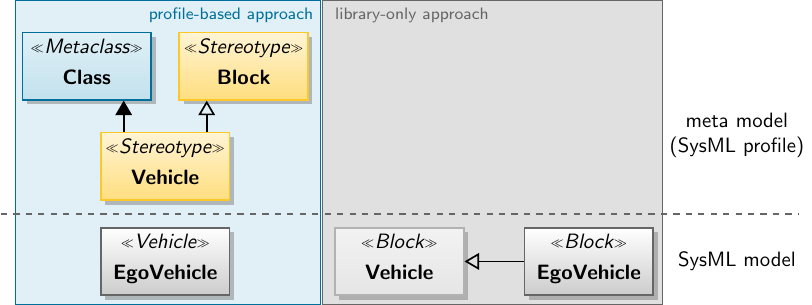}
	\caption{Illustration of library-only vs. profile-based approach for domain-specific modeling in SysML. Left: The domain-specific stereotype \emph{\guillemetleft Vehicle\guillemetright} on the left inherits from the unspecific Stereotype \emph{\guillemetleft Block\guillemetright}. The custom stereotype is used in the model. Right: All model elements are typed with standard stereotype \emph{\guillemetleft Block\guillemetright}, a library component ``Vehicle'' provides a template for a specialized model element ``EgoVehicle''.} %
	\label{fig:library-vs-profile}
\end{figure}

To overcome these challenges, two approaches can be taken (cf. \Cref{fig:library-vs-profile}):
The first is based on so-called ``libraries'': No special SysML profile is created in this case.
Specific model elements can inherit from generic, domain-specific elements, comparable to object-oriented approaches in modern programming languages. 
While the use of such libraries is beneficial for reusing knowledge which has already been captured in specific model elements, the model's stereotypes remain generic and are not tailored toward the application domain.

An approach for such tailoring is described by \textcite{terrasse2006}.
This approach includes the creation of domain-specific meta models.
The authors suggest a three-stage process that shall assist capturing domain-specific expert knowledge and apply this knowledge for system modeling.
According to \textcite{terrasse2006}, this helps to ensure that created models stay consistent to captured knowledge.
The proposed process is built on
\begin{enumerate}[itemsep=0.2em, topsep=0.4em]
	\item capturing domain-specific knowledge in ontologies,
	\item creating domain-specific meta models from the ontologies, and
	\item applying the meta models for creating domain-specific models.
\end{enumerate}

Translated to a SysML-supported MBSE approach, the creation of the meta models requires domain-specific \emph{SysML profiles}.
These profiles provide custom SysML \emph{Stereotypes} for model elements and their respective relations.
This process is also followed by current MBSE-heavy architecture frameworks such as the Unified Architecture Framework (UAF, \citenormy{iso19540}):
The creators of the UAF provide a plethora of concept models, modeled as ontologies, to capture relevant knowledge about the design of enterprise architectures.
Based on these ontologies, UAF defines the ``Unified Architecture Framework Modeling Language'' (UAFML) \parencite{uaf2022}.

\subsection{From Domain Ontologies to Domain-Specific SysML Profiles}
We adopt this approach, as it has two major advantages compared to a library-only model:
\begin{enumerate}[itemsep=0.2em, topsep=0.4em]
	\item The domain ontologies serve as part of a consistent knowledge-base for the design, development, and assurance process, and
	\item profile elements that are derived from the ontologies are always traceable to the structured expert knowledge.
\end{enumerate}
This allows to establish consistent traceability from the elicitation of expert knowledge at the very beginning of the design process to the actual system design (i.e., architectures and requirements), and possibly also to generated source code and runtime artefacts in later process stages or at system runtime.
\begin{figure}[htbp!]%
	\centering
	\includegraphics[width=.99\textwidth]{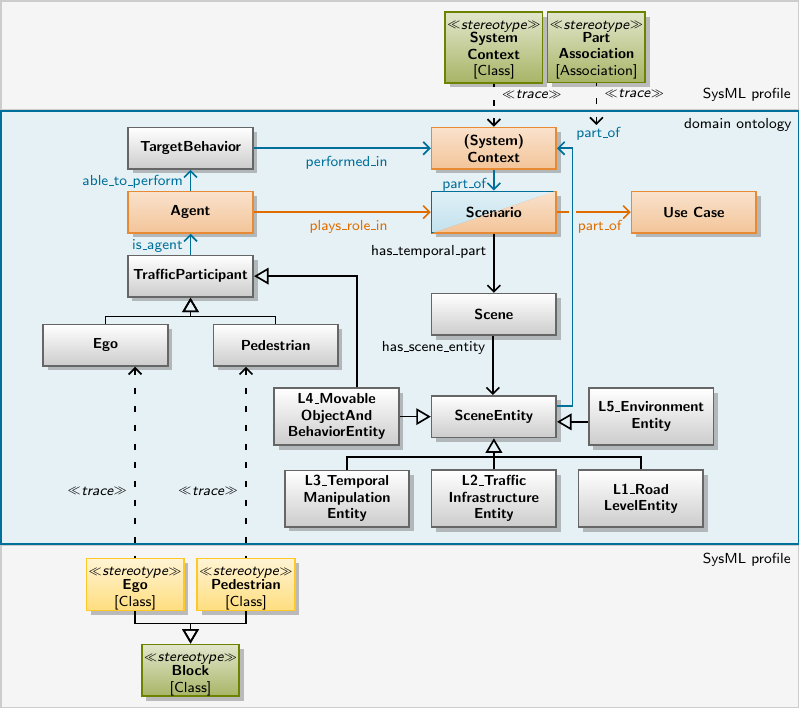}
	\caption{Extract of meta modeling approach for use cases and scenarios at the example of the PEGASUS 6-Layer model \parencite{scholtes2021} (6th layer ommitted): Elements in the domain-specific SysML-profile are traced back to elements of a domain ontology. 
	Grey concepts and relations stem from the automated driving domain, orange elements and relations from the Systems Engineering domain (``Scenarios'' have different notions in both domains), blue arrows depict relations that translate between the domains. 
	Green stereotypes are part of basic SysML 1.4, yellow stereotypes are part of the domain-specific profile.}%
	\label{fig:ad-se-translation}
\end{figure}

In the following, we will give a concrete example for this approach by defining an example scenario which can, e.g., serve as an input for safety analyses.
A simplified excerpt from the created concept models and the domain-specific SysML profile is shown in \Cref{fig:ad-se-translation}.
The ``domain ontology'' contains concepts and relations from the automated driving domain (AD domain, gray boxes, black arrows) and the Systems Engineering domain (SE domain, orange boxes and arrows).
For the automated driving domain (AD-domain), the ontology defines that the \emph{Ego Vehicle} and \emph{Pedestrians} are \emph{Traffic Participants} which are \emph{entities} at Layer 4 of the 6-Layer model. 
Following \textcite{ulbrich2015}, these entities constitute elements of a \emph{Scene}, while a \emph{Scenario} consists of a temporal sequence of \emph{Scenes}.
In the AD domain \emph{and} the SE domain, scenarios can be part of a \emph{Use Case} that summarizes similar scenarios.\footnote{Note that this relation is already defined by \textcite{ulbrich2015} in relation to established Systems Engineering practice.}
An important concept in the SE domain is the \emph{(System) Context} which defines the elements that a System of Interest needs to interact with.
Translated to the AD domain, the \emph{System Context} can be described by the union of all \emph{Scene Entities}, which an automated vehicle interacts with, in a scenario.
Hence, the \emph{System Context} is considered as a \emph{part} of a scenario, whereas \emph{Scene Entities} are part of the \emph{System Context} itself.

In the SE domain, The Unified Architecture Framework introduces the concept of an \emph{Agent} which is an abstract, generalizing concept of an entity that can show their own intentions in a use case (or scenario) by exhibiting behavior.\footnote{In the UAF ontology, agents are implemented by resources which can be systems, people or parts of an organization.\parencite{uaf2022a}}
We add the agent as an intermediate concept that generalizes \emph{Traffic Participants} at Layer 4 of the 6-Layer Model.
In the example, this includes the \emph{Ego} Vehicle and a \emph{Pedestrian}.
\emph{Agents} are \emph{able to perform} \emph{Target Behavior} in a given \emph{Scenario}, that defines a particular \emph{(System) Context}.
This includes, but is not limited to, a technical definition of the \emph{Ego} Vehicle's \emph{Target Behavior}.
This behavior can, e.g., be modeled in terms of required maneuvers to fulfill its mission.\footnote{By modeling the relations in this way, we leave the option of modeling a defined target behavior in a scenario for other traffic participants, e.g. for defining behavior for simulated scenarios.}  

Examples of how the concepts in the domain ontology are translated to elements in a domain-specific SysML profile can be seen above and below the ontology in \Cref{fig:ad-se-translation}.
We introduce custom SysML stereotypes which specialize existing stereotypes in the SysML 1.4 profile.
Examples are the \emph{\guillemetleft Ego\guillemetright} and \emph{\guillemetleft Pedestrian\guillemetright} Stereotypes which specialize the common \emph{\guillemetleft Block\guillemetright} SysML stereotype.
That the corresponding \emph{Scene Entities} can be part of a \emph{(System) Context} is expressed by tracing the SysML \emph{\guillemetleft PartAssociation\guillemetright} to the \emph{part\_of} relation.
The concept of the \emph{(System) Context} is simply traced to the according, already existing, SysML stereotype.

\subsection{Creating Models Based on Domain-Specific SysML Profiles}
\label{sec:models}
With the domain-specific SysML profile established, it is possible to model domain-specific views (cf. \cite{nolte2023, nolte2024}).
For this example, we will focus on modeling contents of the Operational Concept (cf. \Cref{sec:se-to-de}) in terms of use cases, system contexts and scenarios which can be used to analyze the system of interest from the perspective of stakeholder needs.
As discussed, the contents of the Operational Concept are also important artefacts for starting the safety lifecycle according to \citenormy{iso2018} and \citenormy{iso21448}.
By providing parts of a model-based Operational Concept, we introduce the basis for the traceable definition and implementation of safety requirements along the assurance process. 

\begin{figure}[htbp!]
	\centering
        \includegraphics[width=.8\textwidth]{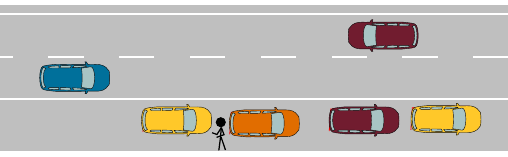}
	\caption{Example scenario that we will use for the argumentation: The ego vehicle passes a row of parked vehicles, a pedestrian is stepping into ego vehicle's driving lane, and is occluded by the parked vehicles.\label{fig:scenario}}
\end{figure}

\begin{figure}[htbp!]
	\centering
    \includegraphics[width=.99\textwidth]{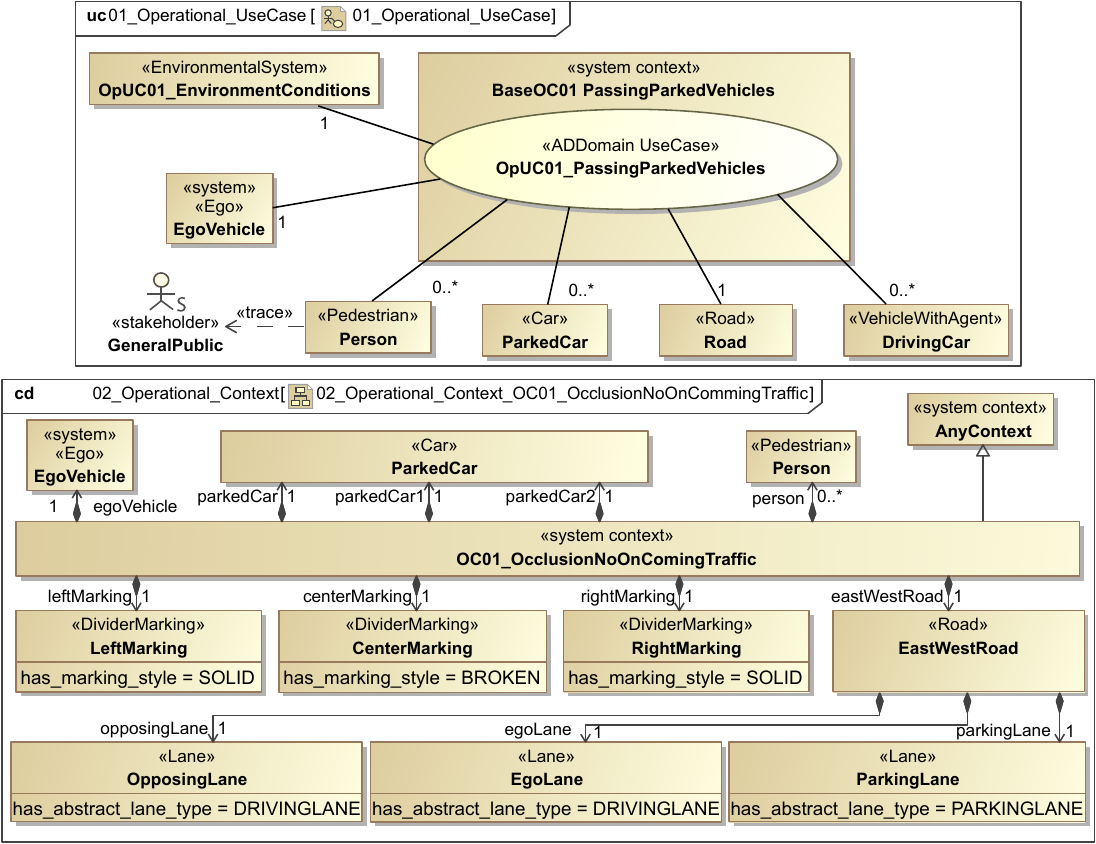}
	\caption{SysML (Operational) Use Case (top) and (Operational) System Context Diagram (bottom) representing the scenario in \Cref{fig:scenario}.\label{fig:use-case}}
\end{figure}

The example is based on the scenario that is shown in \Cref{fig:scenario}:
The \emph{Ego Vehicle} passes a row of parked \emph{Vehicles}.
A \emph{Pedestrian} is about to step onto the street between the first and second parked \emph{Vehicle}.
The scenario can be modeled in SysML as part of a \emph{Use Case} ``Passing Parked Vehicles'' (\Cref{fig:use-case}, top) which subsumes similar scenarios.
While use cases give a general impression with which entities a system of interest has to interact (\emph{Actors}, \cite{friedenthal2015}), the \emph{Context Diagram} (\Cref{fig:use-case}, bottom) provides additional detail per scenario:
It can capture how many actors of a particular kind are present in a scenario; the actors can be decomposed further for providing additional details.
Note that the actors carry domain-specific stereotypes, which according to \textcite{terrasse2006}, can enable easier communication about the related models between stakeholders. 

In the particular example, it can be seen that the ego vehicle needs to interact with the parked vehicles, the road markings, and the pedestrian.
For a full specification of all relevant scene entities in the scenario, the road is decomposed into its lanes and the parking lane. 
The corresponding lane markings are direct parts of the system context due to the concepts in the six layer domain ontology.\footnote{As a context diagram typically only shows directed \emph{part of} relationships (directed association with filled diamond), the relation between divider markings and lanes (\emph{Lane} ``has neighbor'' \emph{Divider}) is not explicitly displayed.}
Individual elements of the use case (or system context) can also be traced directly to stakeholders who might be affected by the automated driving system in a given scenario.
In the given example, the person in \Cref{fig:use-case} (top) is traced to the stakeholder group representing the general public.

\begin{figure}
	\centering
        \includegraphics[width=.99\textwidth]{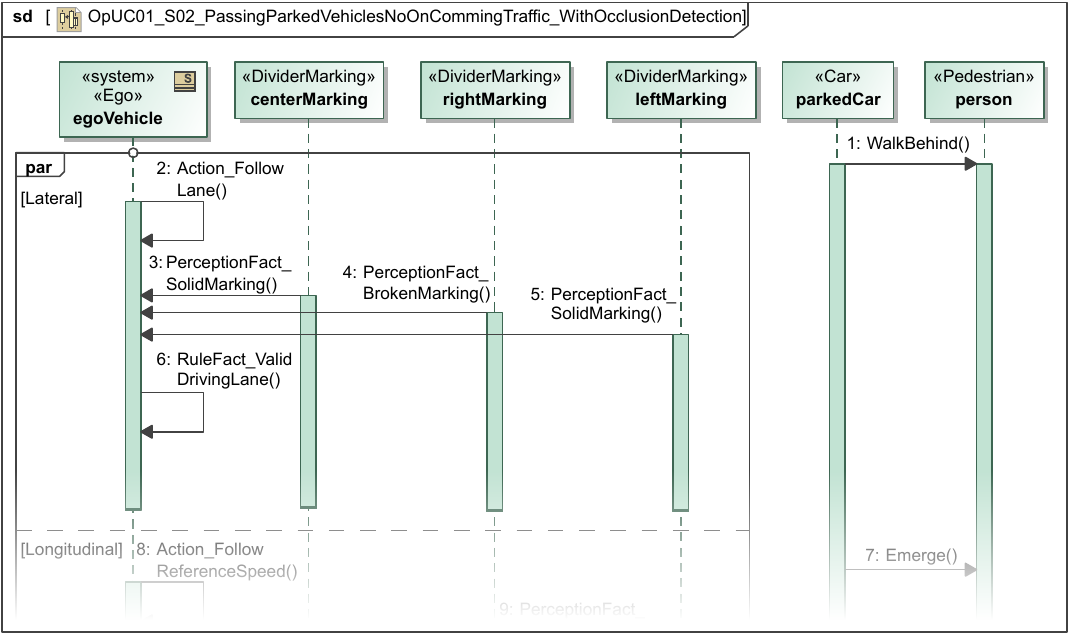}
	\caption{Excerpt from SysML sequence diagram that defines the ego vehicle's behavior in the scenario.\label{fig:sequence-chart}}
\end{figure}

With \emph{Sequence Diagrams}, SysML provides a diagram type which can be used to create SysML representations of functional \parencite{menzel2018} or abstract \parencite{neurohr2021} scenarios (cf. \Cref{fig:sequence-chart}), based on the system context.
The resulting sequence chart depicts how the automated vehicle needs to interact with the elements of the system context to fulfill its mission.

\section{Conclusion \& Future Work}
The models shown in \Cref{sec:models} give an example how parts of an Operational Concept for an automated driving system can be modeled in SysML.
As discussed in \Cref{sec:se-to-de}, a well-defined Operational Concept can assist in merging general Systems Engineering Processes with existing automotive safety processes according to \citenormy{iso2018} and \citenormy{iso21448}.
In this regard, the presented models can provide a semiformal option to model the artefacts required for initiating hazard and risk assessments that are compatible with both standards, which require
\begin{itemize}[itemsep=0.1em, topsep=0.4em]
	\item use cases,
	\item (operational) scenarios,
	\item interactions of the automated driving system / item with its environment, and
	\item traceable relations to stakeholder needs and requirements. 
\end{itemize}

\begin{figure}
	\centering
        \includegraphics[width=.99\textwidth]{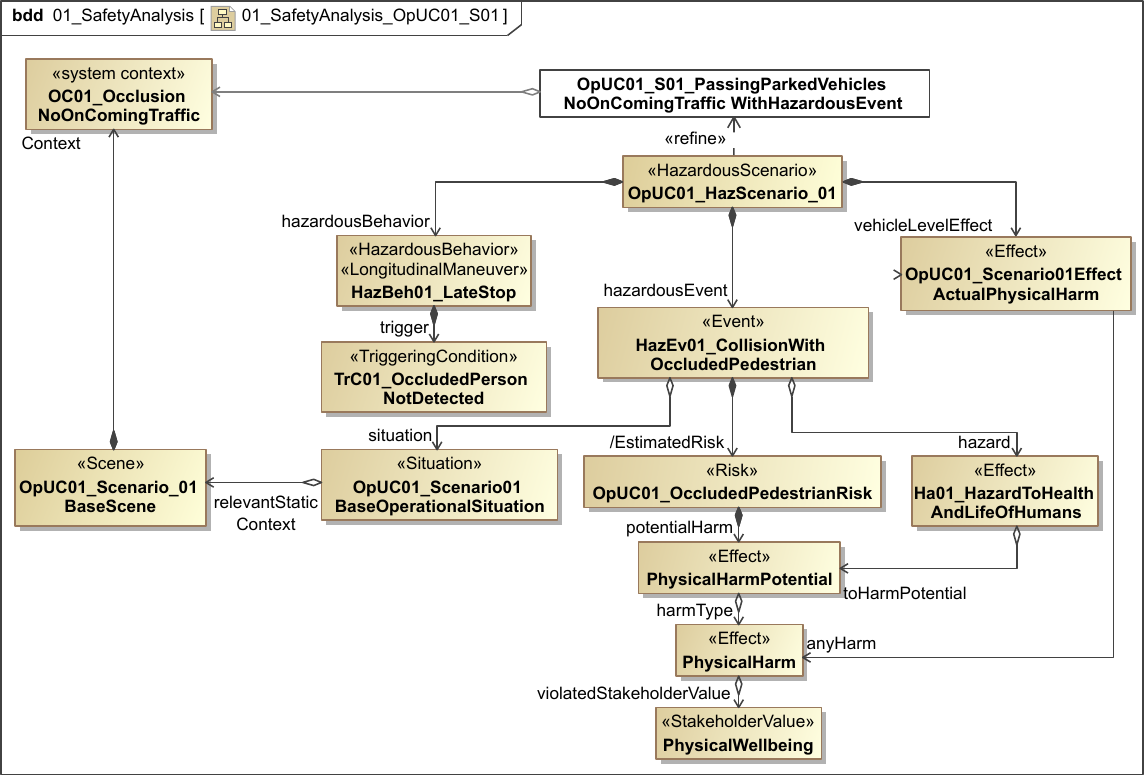}
	\caption{Example model-based safety analysis and modeled safety requirements based on the meta models and models from \Cref{sec:basic-process}. \label{fig:safety-analysis-requirements}}
\end{figure}

The application of AD-domain-specific knowledge and terminology for the creation of the models is an option to provide better communication about the contents of the models.
Last but not least, the SysML models can also be used as a basis to conduct model-based safety analysis to support the generation of traceable safety concepts (\Cref{fig:safety-analysis-requirements}).

While the discussed approach shows great potential to support the traceable design of safe automated driving systems, model-based approaches suffer similar scalability issues as all scenario-based approaches:
Due to the complexity of the real, open world, it is impossible to capture all relevant scenarios at design time.
Efficient design approaches must hence strike a balance between deriving context-specific and context-agnostic requirements.

However, as e.g. \textcite{koopman2024} and \textcite{favaro2023} argue, net risk arguments\footnote{I.e., ``The system is $x$ times better than a human driver in its ODD.''} will always require support from edge- or corner-case-based (and hence scenario-based) arguments that an automated driving system is ``free from unreasonable risk''.
In this respect, the scenario-based hazard and risk assessments which can be supported by the presented approach will always be required.
Model-based consistency and traceability of the created artefacts can greatly assist consistent assurance processes.

Future work will focus on simplifying and automating the modeling process for non Systems Engineers:
Even with the domain-specific models, the manual creation of the discussed diagrams can still be tedious, specifically for domain experts who have little to no MBSE experience.
We assume that simplifying the modeling process can help the acceptance of MBSE-methods in practice.
With the existing meta models as a ground truth that ensures consistency, the creation of what-you-see-is-what-you-get tools that enable a guided modeling process seems promising. 
As the entire approach is based on formalized expert knowledge, it also seems promising to explore how the model-generation process can be automated by combining the existing knowledge with ODD \cite{irvine2021} and scenario description languages \citep{bock2019}.

\section*{Acknowledgement}
We would like to thank our former partners in the project VVMethods (funded by the German Federal Ministry for Economic Affairs and Climate Action, 2019--2023) for fruitful discussions that led to the conclusions presented in this paper.
We would like to thank the German Federal Ministry for Economic Affairs and Climate Action for funding parts of this work in the context of the Project ``Automatisierter Transport zwischen Logistikzentren auf Schnellstraßen im Level 4 (ATLAS-L4)''.

\printbibliography

\end{document}

%% file: tikzstyles.tex
\tikzset
{
	set stroke arrow/.code={%
		\pgfqkeys{/tikz/stroke arrow}{#1}},
	set stroke arrow={%
		stroke width/.initial=.1em,
		stroke color/.initial=black,
		arrow tip start/.initial=,
		arrow tip end/.initial=,
		tip length/.initial=\the\pgflinewidth,
		tip width/.initial=1.5*\the\pgflinewidth,
		new tip width/.initial=1,
		new tip length/.initial=1,
		shorten front/.initial=0,
		shorten back/.initial=0
	},
	do calc/.code={%
		\pgfmathsetmacro{\b}{\pgfkeysvalueof{/tikz/stroke arrow/tip width}/2}
		\pgfmathsetmacro{\a}{\pgfkeysvalueof{/tikz/stroke arrow/tip length}}
		\pgfmathsetmacro{\c}{sqrt(\a*\a + \b*\b)}
		\pgfmathsetmacro{\hOld}{\a * \b / \c}
		\pgfmathsetmacro{\newwidth}{2*(\hOld+1.5*\pgfkeysvalueof{/tikz/stroke arrow/stroke width})*\c / \a}
		\pgfmathsetmacro{\newlength}{(\hOld+1.5*\pgfkeysvalueof{/tikz/stroke arrow/stroke width})*\c / \b}
		\pgfkeyssetvalue{/tikz/stroke arrow/new tip width}{\newwidth}
		\pgfkeyssetvalue{/tikz/stroke arrow/new tip length}{\newlength}
	},
	stroke arrow/.search also={/tikz},
	stroke arrow/.style= {
		set stroke arrow={#1},
		do calc,
		shorten <= \pgfkeysvalueof{/tikz/stroke arrow/shorten front},
		shorten >=\pgfkeysvalueof{/tikz/stroke arrow/shorten back},
		{\pgfkeysvalueof{/tikz/stroke arrow/arrow tip start}[width=\pgfkeysvalueof{/tikz/stroke arrow/tip width}, length=\pgfkeysvalueof{/tikz/stroke arrow/tip length}]}-{\pgfkeysvalueof{/tikz/stroke arrow/arrow tip end}[width=\pgfkeysvalueof{/tikz/stroke arrow/tip width}, length=\pgfkeysvalueof{/tikz/stroke arrow/tip length}]},
		preaction={%
			draw=\pgfkeysvalueof{/tikz/stroke arrow/stroke color},
			shorten <=\pgfkeysvalueof{/tikz/stroke arrow/shorten front}-1.6*\pgfkeysvalueof{/tikz/stroke arrow/stroke width},
			shorten >= \pgfkeysvalueof{/tikz/stroke arrow/shorten back}-1.6*\pgfkeysvalueof{/tikz/stroke arrow/stroke width},			
			line width=\the\pgflinewidth + \pgfkeysvalueof{/tikz/stroke arrow/stroke width},
			{\pgfkeysvalueof{/tikz/stroke arrow/arrow tip start}[width=\pgfkeysvalueof{/tikz/stroke arrow/new tip width}, length=\pgfkeysvalueof{/tikz/stroke arrow/new tip length}]}-{\pgfkeysvalueof{/tikz/stroke arrow/arrow tip end}[width=\pgfkeysvalueof{/tikz/stroke arrow/new tip width}, length=\pgfkeysvalueof{/tikz/stroke arrow/new tip length}]},
		},
	},
}

\tikzset
{
	set road/.code={\pgfqkeys{/tikz/road}{#1}},
	set road={%
		marking width/.initial=\the\pgflinewidth,%
		lane width/.initial=5*\the\pgflinewidth,%
		dash length/.initial=\pgfkeysvalueof{/tikz/road/marking width},%
		color/.initial=lightgray,%
		opacity/.initial=1,%
		curb/.initial=0.2*\pgfkeysvalueof{/tikz/road/lane width},%
		curb left/.initial=\pgfkeysvalueof{/tikz/road/curb},%
		curb right/.initial=\pgfkeysvalueof{/tikz/road/curb},%
		proxy marking right/.style={solid, white, line width=\pgfkeysvalueof{/tikz/road/marking width}},
		marking right/.style={proxy marking right/.style={#1}},
		proxy marking left/.style={solid, white, line width=\pgfkeysvalueof{/tikz/road/marking width}},
		marking left/.style={proxy marking left/.style={#1}},
		proxy marking center/.style={white, dash pattern=on \pgfkeysvalueof{/tikz/road/dash length} off \pgfkeysvalueof{/tikz/road/dash length}, line width=\pgfkeysvalueof{/tikz/road/marking width}},
		marking center/.style={proxy marking center/.style={#1}}
	},
	road/.search also={/tikz},
	road/.style= {
		draw=none,
		set road={#1},
		preaction={
			line width=2*\pgfkeysvalueof{/tikz/road/lane width},
			draw=\pgfkeysvalueof{/tikz/road/color},
			opacity=\pgfkeysvalueof{/tikz/road/opacity},
			shorten >= 0.1ex
		},
		postaction={decorate},
		decoration={
			markings,
			mark=at position 1ex with
			{
			},
			mark=between positions 0 and 1 step 0.5ex with
			{
				\pgfkeysgetvalue{/pgf/decoration/mark info/sequence number}{\j}
				\pgfmathtruncatemacro{\i}{\j-1}
				\coordinate (A\i) at (0,\pgfkeysvalueof{/tikz/road/lane width});
				\coordinate (B\i) at (0,0);
				\coordinate (C\i) at (0,-\pgfkeysvalueof{/tikz/road/lane width});
			},
			mark=at position 0.5 with
			{
				\pgfkeysgetvalue{/pgf/decoration/mark info/sequence number}{\j}
				\pgfmathtruncatemacro{\marks}{\j-2}
				\pgfmathparse{\pgfkeysvalueof{/tikz/road/curb right}}
				\ifnum 0=\pgfmathresult
				\else
				\draw[solid,opacity=\pgfkeysvalueof{/tikz/road/opacity},\pgfkeysvalueof{/tikz/road/color}, line width=\pgfkeysvalueof{/tikz/road/curb right}] ($(C\marks.center)+(0,-0.5*\pgfkeysvalueof{/tikz/road/curb right}-0.4*\pgfkeysvalueof{/tikz/road/marking width})$)
				foreach \i in {\marks,...,1}
				{ -- ($(C\i.center)+(0,-0.5*\pgfkeysvalueof{/tikz/road/curb right}-0.4*\pgfkeysvalueof{/tikz/road/marking width})$) } -- ($(C1.center)+(0,-0.5*\pgfkeysvalueof{/tikz/road/curb right}-0.4*\pgfkeysvalueof{/tikz/road/marking width})$);
				\fi
				\pgfmathparse{\pgfkeysvalueof{/tikz/road/curb left}}
				\ifnum 0=\pgfmathresult
				\else	            		
				\draw[solid,opacity=\pgfkeysvalueof{/tikz/road/opacity},\pgfkeysvalueof{/tikz/road/color}, line width=\pgfkeysvalueof{/tikz/road/curb left}] ($(A\marks.center)-(0,-0.5*\pgfkeysvalueof{/tikz/road/curb left}-0.4*\pgfkeysvalueof{/tikz/road/marking width})$)
				foreach \i in {\marks,...,1}
				{ -- ($(A\i.center)-(0,-0.5*\pgfkeysvalueof{/tikz/road/curb left}-0.4*\pgfkeysvalueof{/tikz/road/marking width})$) } -- ($(A1.center)-(0,-0.5*\pgfkeysvalueof{/tikz/road/curb left}-0.4*\pgfkeysvalueof{/tikz/road/marking width})$);
				\fi
				\draw[road/proxy marking right] (C\marks.center)
				foreach \i in {\marks,...,1}
				{ -- (C\i.center) } -- (C1.center);
				\draw[road/proxy marking left] (A\marks.center)
				foreach \i in {\marks,...,1}
				{ -- (A\i.center) } -- (A1.center);
				\draw[road/proxy marking center] (B\marks.center)
				foreach \i in {\marks,...,1}
				{ -- (B\i.center) } -- (B1.center);
			}
		},
	},
}
\makeatletter
\pgfdeclaredecoration{free hand}{start}
{
	\state{start}[width = +0pt,
	next state=step,
	persistent precomputation = \pgfdecoratepathhascornerstrue]{}
	\state{step}[auto end on length    = 3pt,
	auto corner on length = 3pt,               
	width=+2pt]
	{
		\pgfpathlineto{
			\pgfpointadd
			{\pgfpoint{2pt}{0pt}}
			{\pgfpoint{rand*0.15pt}{rand*0.15pt}}
		}
	}
	\state{final}
	{}
}
\makeatother
\tikzset{
	double yellow/.style = {%
		double=lightgray,
		double distance=0.8*\mylinewidth,
		line width=\mylinewidth,
		draw=tuYellow
	}
}
\tikzset{
	dashed marking/.style args={#1 dashlength #2}{%
		#1,
		dash pattern=on #2 off #2
	},
	dashed centerline german/.style 2 args={%
		dashed marking={solid, white, line width=#1 dashlength #2}
	}
}
\tikzset{%
	place nodes/.style={%
		postaction={%
			decorate,%
			decoration={%
				markings,%
				mark=between positions 0 and 1 step 0.125 with {%
					\node [] (a\pgfkeysvalueof{/pgf/decoration/mark info/sequence number}) {};%
				},%
			}%
		}%
	}%
}%

\pgfdeclaredecoration{triangle}{start}{
	\state{start}[width=0.99\pgfdecoratedinputsegmentremainingdistance,next state=up from center]
	{\pgfpathlineto{\pgfpointorigin}}
	\state{up from center}[next state=do nothing]
	{
		\pgfpathlineto{\pgfqpoint{\pgfdecoratedinputsegmentremainingdistance}{\pgfdecorationsegmentamplitude}}
		\pgfpathlineto{\pgfqpoint{\pgfdecoratedinputsegmentremainingdistance}{-\pgfdecorationsegmentamplitude}}
		\pgfpathlineto{\pgfpointdecoratedpathfirst}
	}
	\state{do nothing}[width=\pgfdecorationsegmentlength,next state=do nothing]{
		\pgfpathlineto{\pgfpointdecoratedinputsegmentfirst}
		\pgfpathmoveto{\pgfpointdecoratedinputsegmentlast}
	}
}

\tikzset{%
	stop sign/.style 2 args={%
		rounded corners=#1*0.2pt,%
		xshift=#1*\lanewidth,%
		shape=regular polygon,%
		line width=#1*0.8pt,%
		regular polygon sides=8,%
		fill=tuRed,%
		draw=white,%
		inner sep=#1*-0.2pt,%
		label={%
			[anchor=center, 
			yshift=#1*-0.7em, 
			text=white, 
			font=#2\sffamily\bfseries]
			\scalebox{0.5}[1.2]{STOP}},%
		minimum width=#1*1.4em,%
		append after command={\pgfextra%
			\draw[line width=#1*0.3pt, black, rounded corners=#1*0.1pt]%
			($(\tikzlastnode.center)!1.01!(\tikzlastnode.corner 1)$) --%
			($(\tikzlastnode.center)!1.01!(\tikzlastnode.corner 2)$) --%
			($(\tikzlastnode.center)!1.01!(\tikzlastnode.corner 3)$) --%
			($(\tikzlastnode.center)!1.01!(\tikzlastnode.corner 4)$) --%
			($(\tikzlastnode.center)!1.01!(\tikzlastnode.corner 5)$) --%
			($(\tikzlastnode.center)!1.01!(\tikzlastnode.corner 6)$) --%
			($(\tikzlastnode.center)!1.01!(\tikzlastnode.corner 7)$) --%
			($(\tikzlastnode.center)!1.01!(\tikzlastnode.corner 8)$) -- cycle;%
			\endpgfextra%
		}%
	}%
}
\tikzset{%
	psm fact/.style={%
		ellipse,
		drop shadow,
		shadedGray,
		align=center,
		text width=6em,
		minimum height=3em
	},
	detection fact/.style={
		psm fact,
		shadedBlueLight
	},
	action fact/.style={
		psm fact,
		shadedRedLight
	},	
	rule fact/.style={
		psm fact,
		shadedGreenLight
	}	
}

\tikzset{
	triangle path/.style={decoration={triangle,amplitude=#1}, decorate},
	triangle path/.default=1ex}

\tikzset{%
	capability block/.style={
		minimum width=5.5em,
		minimum height=4.5em,
		align=center,
	},
	indicator/.style={
		capability block,
		shadedGray,
	},
	capability fact/.style={%
		capability block,
		shadedBlueLight,
		rounded corners=0.5em,
	},
	myCapability/.style={%
	trapezium,
	trapezium left angle=80,
	trapezium right angle=-80,
	minimum width=4em,
	minimum height=4.2em,
	text width=7.5em,
	align=center,
	inner sep=0.5em
},
	knowledge cap/.style={%
		myCapability,
		shadedGrayLight,
	},
	scenario cap/.style={%
		myCapability,
		shadedGrayDark,
	},
	abstract cap/.style={%
		scenario cap,
		dashed
	},
	connect/.style={%
		thick,
		-{[round]Stealth},
		rounded corners=2pt
	}	
}

\tikzstyle{shadedRedLight} = [top color=white!90!tuRed, bottom color=tuRed!60!white, draw=tuRed80, thick]%
\tikzstyle{shadedBlue} = [top color=tuDarkBlue20, bottom color=tuDarkBlue40, draw=tuDarkBlue80, thick]%
\tikzstyle{shadedBlueLight} = [top color=tuLightBlue20, bottom color=tuLightBlue40, draw=tuBlue, thick]%
\tikzstyle{shadedRed} = [top color=tuRed20, bottom color=tuRed40, draw=tuRed80, thick]%
\tikzstyle{shadedOrange}  = [top color=tuOrange20, bottom color=tuOrange60,  draw=tuOrange100, thick]%
\tikzstyle{shadedOrangeDark}  = [top color=tuOrange40!99!black, bottom color=tuOrange80!99!black,  draw=tuOrange100!99!black, thick]%
\tikzstyle{shadedOrangeLight}  = [top color=tuOrange20,bottom color=tuOrange40, draw=tuOrange80, thick]%
\tikzstyle{shadedGreen}  = [top color=tuGreen20, bottom color=tuGreen60,  draw=tuGreen100, thick]%
\tikzstyle{shadedGreenLight}  = [top color=tuLightGreen20, bottom color=tuLightGreen60,  draw=tuLightGreen100, thick]%
\tikzstyle{shadedYellow} = [top color=tuYellow20, bottom color=tuYellow60,  draw=tuYellow100, thick]%
\tikzstyle{shadedGray} = [top color=white, bottom color=tuGray20, draw=tuGray60, thick]%
\tikzstyle{shadedGrayLight} = [top color=tuBlack!5, bottom color=tuBlack!10, draw=tuBlack!30, thick]%
\tikzstyle{shadedGrayDark} = [top color=tuBlack!20, bottom color=tuBlack!40, draw=tuBlack!60, thick]%
\tikzstyle{shadedGrayMedium} = [top color=tuBlack!13, bottom color=tuBlack!25, draw=gray, thick]%
\tikzstyle{shadedGrayMediumLight} = [top color=tuBlack!10, bottom color=tuBlack!18, draw=gray, thick]%
\tikzstyle{shadedGrayMediumDark} = [top color=tuBlack!15, bottom color=tuBlack!32, draw=darkgray, thick]%
\tikzstyle{shadedBlueMedium} = [top color=tuLightBlue40, bottom color=tuBlue40, draw=tuBlue, thick]%
\tikzstyle{shadedViolet} = [top color=white!97!tuViolet, bottom color=tuViolet!30!white, draw=tuViolet80, thick]%
\tikzstyle{shadow} = [drop shadow={opacity=.5,shadow xshift=.3ex,shadow yshift=-.3ex}]%
\tikzstyle{triangle} = [isosceles triangle,isosceles triangle stretches]%
\tikzstyle{label-it} = [font=\itshape]%
\tikzstyle{block} = [draw, shadedBlue, rectangle, rounded corners, minimum height=2em, minimum width=5em]%
\tikzstyle{smallblock} = [draw, shadedBlue, rectangle, rounded corners, minimum height=1em, minimum width=2em,shadow]%
\tikzstyle{outerblock} = [draw, shadedGrayLight, draw=tuGray80, rectangle, rounded corners, minimum height=2em, minimum width=5em,shadow]%
\tikzstyle{bubble} = [fill=black,shadow,circle,draw=black,inner sep=0pt,minimum size=5pt]%
\tikzstyle{memory} = [cylinder, shape border rotate=90, aspect=.4, shadedGrayLight, minimum width=5em, shadow]%
\tikzstyle{inheritArrow} = [-open triangle 60,thick]%
\tikzstyle{kompArrow}    = [diamond-,thick]%
\tikzstyle{flowDecision} = [diamond, draw, shadedRed, text badly centered, inner sep=0pt,shadow]%
\tikzstyle{flowBlock} = [rectangle, draw, shadedBlue, text centered, rounded corners, minimum height=2em,shadow]%
\tikzstyle{bgBox} = [rectangle, draw, shadedGrayLight, text centered, rounded corners=3mm, shadow, inner sep=10pt]%
\tikzstyle{blockarrow} = [draw, thick, single arrow, minimum height=3em]%

\tikzstyle {archblock} = [outerblock, minimum height=4em, align=center, minimum width=12em, font=\sffamily]
\tikzstyle {slim} = [minimum width=6em]
\tikzstyle {verticalblock} = [archblock, rotate = 90, archblock, minimum width=16.5em, minimum height=2em, on grid]
\tikzstyle {sensors} = [rectangle split parts=2, rectangle split horizontal]

\tikzstyle {sup} = [yshift=0.5cm]
\tikzstyle {sdown} = [yshift=-0.5cm]
\tikzstyle {sleft} = [xshift=-0.5cm]
\tikzstyle {sright} = [xshift=0.5cm]

\tikzstyle {arrow} = [very thick, -{Stealth[round]}]

\tikzstyle {seperator} = [thick, dashed, lightgray]
\tikzstyle {arrow} = [-stealth', thick]
\tikzstyle {rarrow} = [arrow, stealth'-]

\tikzstyle{maneuver area} = [fill=tuDarkBlue40, fill opacity=0.6]
\tikzstyle{blue car} = [vehicle, draw=darkgray!20!black, fill=tuBlue]
\tikzstyle{red car} = [vehicle, draw=darkgray!20!black, fill=tuRed100]
\tikzstyle{orange car} = [vehicle, draw=darkgray!20!black, fill=tuOrange]
\tikzstyle{yellow car} = [vehicle, draw=darkgray!20!black, fill=tuYellow]
\tikzstyle{green car} = [vehicle, draw=darkgray!20!black, fill=tuGreen]
\tikzstyle{target pose} = [opacity=0.7, draw=white]

\makeatletter
\tikzset{reset preactions/.code={\def\tikz@preactions{}}}
\makeatother

\tikzstyle {disable} = [reset preactions, draw=none, fill=none, top color=white, bottom color=white]
\tikzstyle {header} = [disable, font=\sffamily]

\tikzset{
	fit label/.style={yshift={(height("#1")+4pt)/2},
	inner ysep={(height("#1")+8pt)/2},
	label={[anchor=north,font=\itshape]north:#1 \pgfkeysvalueof{/pgf/inner xsep}}},
	free hand/.style={
			decorate, thick, draw,
			decoration={free hand}
	},  
}

\tikzset{
	double arrow/.style args={#1 #2 colored by #3 and #4}{
		-{#1[round]},line width=#2,#3, 
		postaction={draw,-{#1[round, scale=1.095]},#4,line width=(#2)/2,
			shorten <=0.6*(#2)/2,shorten >=0.6*(#2)/2}, 
	}
}

\tikzset{mysplit/.style={rectangle split, rectangle split parts=2, rectangle split draw splits=false, inner sep=2.2ex,
		rectangle split horizontal,minimum width=5.5em},
	textstyle/.style={text height=1.5ex,text depth=.25ex}}

\tikzumlset{class width=18ex}

\tikzumlset{draw element=gray!95!black}
\tikzumlset{font=\small}
\tikzstyle{tikzuml simpleclass style}=[shadedGray, shadow, minimum width=5em, align=center, font=\textbf]
\pgfkeys{%
	/tikzuml/relation/style/.style={#1, font=\small}
}

\tikzset{
	TLeft/.style={label={[anchor=east, xshift=1ex]left:$\blacktriangleleft$}},
	TRight/.style={label={[anchor=west, xshift=-1ex]right:\strut$\blacktriangleright$}},
	TUp/.style={label={[anchor=south, yshift=-1ex]above:$\blacktriangle$}},
        TUpR/.style={label={[anchor=west, xshift=-1.5ex]right:$\blacktriangle$}},
        TUpL/.style={label={[anchor=east, xshift=1.5ex]left:$\blacktriangle$}},
	TDown/.style={label={[anchor=north, yshift=1ex]below:$\blacktriangledown$}},
        TDownR/.style={label={[anchor=west, xshift=-1.5ex]right:$\blacktriangledown$}},
        TDownL/.style={label={[anchor=east, xshift=1.5ex]left:$\blacktriangledown$}},
	closer pos/.style={},
	VP/.style={{archblock, font=\normalfont}}
}

\pgfdeclarepatternformonly{swnestripes}{\pgfpoint{0cm}{0cm}}{\pgfpoint{1cm}{1cm}}{\pgfpoint{1cm}{1cm}}
{
	\foreach \i in {0.1, 0.3,...,0.9}
	{
		\pgfpathmoveto{\pgfpoint{\i cm}{0cm}}
		\pgfpathlineto{\pgfpoint{1cm}{1cm - \i cm}}
		\pgfpathlineto{\pgfpoint{1cm}{1cm - \i cm + 0.1cm}}
		\pgfpathlineto{\pgfpoint{\i cm- 0.1cm}{0cm}}
		\pgfpathclose%
		\pgfusepath{fill}
		\pgfpathmoveto{\pgfpoint{0cm}{\i cm}}
		\pgfpathlineto{\pgfpoint{1cm - \i cm}{1cm}}
		\pgfpathlineto{\pgfpoint{1cm - \i cm- 0.1cm}{1cm}}
		\pgfpathlineto{\pgfpoint{0cm}{\i cm+ 0.1cm}}
		\pgfpathclose%
		\pgfusepath{fill}
	}
}
\pgfdeclarepatternformonly{swneStripes}{\pgfpoint{0cm}{0cm}}{\pgfpoint{1cm}{1cm}}{\pgfpoint{1cm}{1cm}}
{
	\foreach \i in {0.1, 0.3,...,0.9}
	{
		\pgfpathmoveto{\pgfpoint{\i cm}{0cm}}
		\pgfpathlineto{\pgfpoint{1cm}{1cm - \i cm}}
		\pgfpathlineto{\pgfpoint{1cm}{1cm - \i cm - 0.1cm}}
		\pgfpathlineto{\pgfpoint{\i cm+ 0.1cm}{0cm}}
		\pgfpathclose%
		\pgfusepath{fill}
		\pgfpathmoveto{\pgfpoint{0cm}{\i cm}}
		\pgfpathlineto{\pgfpoint{1cm - \i cm}{1cm}}
		\pgfpathlineto{\pgfpoint{1cm - \i cm+ 0.1cm}{1cm}}
		\pgfpathlineto{\pgfpoint{0cm}{\i cm- 0.1cm}}
		\pgfpathclose%
		\pgfusepath{fill}
	}
}
\pgfdeclarepatternformonly{senwstripes}{\pgfpoint{0cm}{0cm}}{\pgfpoint{1cm}{1cm}}{\pgfpoint{1cm}{1cm}}
{
	\foreach \i in {0.1, 0.3,...,0.9}
	{
		\pgfpathmoveto{\pgfpoint{0cm}{\i cm}}
		\pgfpathlineto{\pgfpoint{0cm}{\i cm+ 0.1cm}}
		\pgfpathlineto{\pgfpoint{\i cm+ 0.1cm}{0cm}}
		\pgfpathlineto{\pgfpoint{\i cm}{0cm}}
		\pgfpathclose%
		\pgfusepath{fill}
		\pgfpathmoveto{\pgfpoint{1cm}{\i cm}}
		\pgfpathlineto{\pgfpoint{\i cm}{1cm}}
		\pgfpathlineto{\pgfpoint{\i cm + 0.1cm}{1cm}}
		\pgfpathlineto{\pgfpoint{1cm}{\i cm+ 0.1cm}}
		\pgfpathclose%
		\pgfusepath{fill}
	}
}
\pgfdeclarepatternformonly{senwStripes}{\pgfpoint{0cm}{0cm}}{\pgfpoint{1cm}{1cm}}{\pgfpoint{1cm}{1cm}}
{
	\foreach \i in {0.0, 0.2,...,0.8}
	{
		\pgfpathmoveto{\pgfpoint{\i cm}{0cm}}
		\pgfpathlineto{\pgfpoint{0cm}{\i cm }}
		\pgfpathlineto{\pgfpoint{0cm}{\i cm+ 0.1cm}}
		\pgfpathlineto{\pgfpoint{\i cm+ 0.1cm}{0cm}}
		\pgfpathclose%
		\pgfusepath{fill}
		\pgfpathmoveto{\pgfpoint{1cm}{\i cm}}
		\pgfpathlineto{\pgfpoint{\i cm}{1cm}}
		\pgfpathlineto{\pgfpoint{\i cm+ 0.1cm}{1cm}}
		\pgfpathlineto{\pgfpoint{1cm}{\i cm + 0.1cm}}
		\pgfpathclose%
		\pgfusepath{fill}
	}
}

\newcounter{coordinateindex}

\tikzset{
	put coordinates/.style={
		initialize counter/.code={
			\setcounter{coordinateindex}{0}
		},
		initialize counter,
		decoration={
			show path construction,
			moveto code={
				\stepcounter{coordinateindex}
				\coordinate (#1-\thecoordinateindex) at (\tikzinputsegmentfirst);
			},
			lineto code={
				\stepcounter{coordinateindex}
				\coordinate (#1-\thecoordinateindex) at (\tikzinputsegmentlast);
			},
			curveto code={
				\stepcounter{coordinateindex}
				\coordinate (#1-\thecoordinateindex) at (\tikzinputsegmentlast);
			},
			closepath code={
				\stepcounter{coordinateindex}
				\coordinate (#1-\thecoordinateindex) at (\tikzinputsegmentlast);
			},
		},
		postaction={decorate}
	},
	put coordinates/.default=coordinate
}

\newcommand{\PrepareOcclusion}[2]%
{%
	\path[name path=AB] %
	\foreach \i in {1,...,\thecoordinateindex}{%
		(#1) -- ($(#1)!100cm!(#2-\i)$)%
	};%
	\foreach \i  [count=\j] in {1,...,\thecoordinateindex}%
	{%
		\coordinate (#2-shadow-\i) at (#2-\i);%
	}%
	\pgfresetboundingbox%
	\draw (-20,-20) rectangle (20,20);%
	\path[name path=CBB] (current bounding box.north west) -- (current bounding box.north east) -- (current bounding box.south east) -- (current bounding box.south west) -- cycle;%
	\path[execute at begin node={\global\let\t=\t}, name intersections={name=int-#2, of=AB and CBB, sort by=AB, total=\t}];

	\foreach \i [count=\k from \thecoordinateindex+1] in {\t,...,1}%
	{%
		\def\nodename{#2}
		\global\expandafter\let\csname maxIdx\nodename\endcsname\k
		\coordinate (#2-shadow-\k) at (int-#2-\i);%
		
	}%
	\pgfresetboundingbox%
}%
\newcommand{\DrawOcclusion}[2]%
{%
	\def\maxIdx{\csname maxIdx#1\endcsname}
	\draw[#2](#1-shadow-1)%
	\foreach \x in {2,3,...,\maxIdx}%
	{%
		-- (#1-shadow-\x)%
	} -- cycle;%
}%

\tikzstyle{function} = [shadedGray, thick, font=\bfseries, minimum width=13em,, inner ysep=0.75em, align=center]%
\tikzstyle{port} = [rectangle, shadedGray, minimum width=1.5em, minimum height=1.5em]%
\pgfdeclarearrow{%
	name=Contains,%
	parameters= {\the\pgfarrowlength},%
	setup code={%
		\pgfarrowssettipend{0pt}%
		\pgfarrowssetlineend{-\pgfarrowlength}%
		\pgfarrowlinewidth=\pgflinewidth%
		\pgfarrowssavethe\pgfarrowlength%
	},%
	drawing code={%
		\pgfpathcircle{\pgfpoint{-0.5\pgfarrowlength}{0pt}}{0.5\pgfarrowlength}%
		\pgfpathmoveto{\pgfpoint{-\pgfarrowlength}{0}}%
		\pgfpathlineto{\pgfpoint{0}{0}}%
		\pgfpathmoveto{\pgfpoint{-0.5\pgfarrowlength}{0.5\pgfarrowlength}}%
		\pgfpathlineto{\pgfpoint{-0.5\pgfarrowlength}{-0.5\pgfarrowlength}}%
		\pgfusepathqstroke%
	},%
	defaults = { length = 0.75em }%
}
\tikzset{%
	hw_base/.style={%
		function, %
		label={[anchor=north,#1]\emph{\guillemotleft block\guillemotright}},%
	},%
	hw/.style={%
		hw_base,%
		inner ysep=1.5em, text depth=1.5em%
	},%
	connection/.style args={#1#2}{%
		very thick,%
		postaction={%
			decorate,%
			decoration={%
				markings,%
				mark=at position 0.5 with {\arrow{Triangle[scale=2]}, \node[{#1,align=center, text height=1ex, text depth=0.25ex, font=\small}] {#2};}%
			}%
		},%
	},%
	req/.style args={#1#2#3#4}{%
		label={[anchor=north]\emph{\guillemotleft #1\guillemotright}},%
		align=center,%
		minimum width=7em,
		font=\scriptsize\bfseries,%
		text height=1.5em,%
		append after command={%
			\pgfextra{%
				\path let \p1=($(\tikzlastnode.east) - (\tikzlastnode.west)$) in node[{name=\tikzlastnode-sup, anchor=north, align=left, align=left, minimum width=\x1}] at (\tikzlastnode.south) {Id=``#2''\\Text=``#3''#4};%
				\draw[shadedGray] (\tikzlastnode-sup.north west) -- (\tikzlastnode-sup.north east);%
				\begin{pgfonlayer}{pre main}%
					\node[fit=(\tikzlastnode)(\tikzlastnode-sup), shadedGray,drop shadow, inner sep=0, name=\tikzlastnode-req]{};%
				\end{pgfonlayer}%
			}%
		}%
	},%
	rel label/.style args={#1#2}{%
		postaction={ decorate,%
			decoration={ markings,%
				mark=at position 0.5 with \node[{#1,text height=1ex, text depth=0.25ex, font=\scriptsize}] {#2};%
			}%
		}%
	},%
	inport/.style={%
		port,%
		append after command={%
			\pgfextra{%
				\begin{pgfonlayer}{foreground}%
					\draw[thick,-stealth', shorten <=.4em, shorten >=.4em, gray ] (\tikzlastnode.west) -- (\tikzlastnode.east);%
				\end{pgfonlayer}%
			}%
		}%
	},%
	ioport/.style={%
		port,%
		append after command={%
			\pgfextra{%
				\begin{pgfonlayer}{foreground}%
					\draw[thick,stealth'-stealth', shorten <=.3em, shorten >=.3em, gray ] (\tikzlastnode.west) -- (\tikzlastnode.east);%
				\end{pgfonlayer}%
			}%
		}%
	},%
	outport/.style={%
		port,%
		append after command={%
			\pgfextra{%
				\begin{pgfonlayer}{foreground}%
					\draw[thick,-stealth', shorten <=.3em, shorten >=.3em, gray ] (\tikzlastnode.east) -- (\tikzlastnode.west);%
				\end{pgfonlayer}%
			}%
		}%
	},%
	hw_top_port/.style={%
		hw_base={yshift=-1em},%
		text height=2.75em,%
		inner ysep=0.5em,%
	},%
	derived-req/.style={%
		-{Straight Barb[length=0.5em,width=0.7em]},%
		dashed,%
		thick%
	},%
	contained-req/.style={%
		thick,%
		-Contains%
	},%
	association/.style ={%
		draw,
		thick
	},%
	usecase/.style={
		ellipse,
		shadedGray,
		drop shadow,
		inner sep=1em,
		align=center,
		font=\bfseries
	},%
	diagramStyle/.style={%
		shadedGray, drop shadow, inner ysep=2em, yshift=1em, inner xsep=0.5em
	},
	diagramFrame/.style={%
		thick,gray,fill=none
	}
}%

\tikzset{%
	set class/.code={\pgfqkeys{/tikz/class}{#1}},%
	set class={%
		stereotype/.store in=\myStereoType,%
		class name/.store in=\myClassName,%
		parts/.store in=\myPartNum,%
		stereotype=,%
		class name=myClass,%
		parts=1%
	},%
	class/.search also={/tikz},%
	class/.style={%
		set class={#1},%
		align=center,%
		inner ysep=0.25em,%
		font=\bfseries,%
		shadedGray,drop shadow,%
		node contents={\myClassName},%
		minimum width=10em,
		class/add stereotype/.expand once=\myStereoType%
	},%
	class/.cd,
	add mystereotype/.code={%
		\def\argone{#1}
		\ifx\argone\empty
		\pgfkeysalso{minimum height=2.5em}
		\else
		\pgfkeysalso{
		label={[anchor=north]\emph{\guillemotleft\vphantom{ghA}{#1}\guillemotright}},%
		text width={max(6em, 1em + max(width(\string"#1\string"),width(\string"\myClassName\string"))))},%
		text height=2.2em,%
		text depth=0.25em,%
		}
		\fi
	},
	add stereotype/.style={%
		class/add mystereotype={#1}
	}
}

\tikzset{
	cog/.style={
		draw,
		thick,
		minimum size=0.3cm,
		centerofmass,
		cog rotate=#1
	},
	cog/.default=0,
	frame-arrow/.style={thick, -stealth'},
	vehicle-velocity/.style={frame-arrow, tuBlue},
	s-velocity/.style={frame-arrow, tuOrange},
	wheelforce/.style={frame-arrow, tuRed}
}
\pgfkeyssetvalue{/tikz/cog rotate}{0}%
\tikzset{sFrameInvisible/.style={opacity=0}}%

\tikzset{wheelFrameInvisible/.style={opacity=1}}

\tikzset{trajectoryInvisible/.style={opacity=1}}

\def\centerarc[#1](#2)(#3:#4:#5)
{ \draw[#1] ($(#2)+({#5*cos(#3)},{#5*sin(#3)})$) arc (#3:#4:#5); }

\def\labeledcenterarc[#1](#2)(#3,#4)(#5:#6:#7)
{ \draw[#1] ($(#2)+({#7*cos(#5)},{#7*sin(#5)})$) arc (#5:#6:#7) node[midway,#4] {#3}; }    

\tikzset{wheel/.style={draw=gray, thick, rounded corners=.1em, fill=lightgray!50}}

\makeatletter
\tikzset{%
	reverse path/.style={
		decoration={show path construction,
			reverse path,
			moveto code={\pgfpathmoveto{\pgf@decorate@inputsegment@first}},
			lineto code={\pgfpathlineto{\pgf@decorate@inputsegment@last}},
			curveto code={\pgfpathcurveto{\pgf@decorate@inputsegment@supporta}%
				{\pgf@decorate@inputsegment@supportb}{\pgf@decorate@inputsegment@last}},
			closepath code={\pgfpathclose}
		},
		decorate
	}
}
\makeatother

\tikzset{%
	generalization/.append style={%
		reverse path
	},
	specialization/.append style={%
	reverse path
},	
}

\tikzset{%
	set association/.code={\pgfqkeys{/tikz/association}{#1}},%
	set association={%
		pos1/.store in=\fromPos,%
		pos2/.store in=\toPos,%
		mult1/.store in=\fromMult,%
		mult2/.store in=\toMult,%
		arg1/.store in=\fromArg,%
		arg2/.store in=\toArg,%
		argpos1/.store in=\fromArgPos,%
		argpos2/.store in=\toArgPos,%
		multpos1/.store in=\fromMultPos,%
		multpos2/.store in=\toMultPos,%
		stereotype/.store in=\stereotype,%
		pos/.store in=\sPos,
		pos1=1em,%
		pos2=-1.5em,%
		mult1=,%
		mult2=,%
		arg1=,%
		arg2=,%
		argpos1=,
		argpos2=,
		multpos1=,
		multpos2=,
		stereotype=,%
		pos=0.5%
	},%
	association/.search also={/tikz},%
	association/.style={%
		set association={#1},%
		draw,%
		thick,%
		postaction={decorate},%
		decoration={%
			markings,%
			mark=at position \fromPos-0.02 with%
			{%
				\node (A) {};%
				\pgfgetlastxy{\XCoord}{\YCoord}%
				\global\let\oldX\XCoord%
				\global\let\oldY\YCoord%
			},%
			mark=at position \fromPos with%
			{%
				\node (B){};	%
				\pgfmathsetmacro\lenArg{width("\fromArg")}				
				\pgfmathsetmacro\lenMult{width("\fromMult")}
				\pgfgetlastxy{\xlast}{\ylast}%
				\pgfmathsetmacro{\xDiff}{(abs(\xlast)-abs(\oldX))*100}%
				\pgfmathsetmacro{\yDiff}{(abs(\ylast)-abs(\oldY))*100}%
				\pgfmathsetmacro{\intXDiff}{int((\xlast-\oldX)*100)}
				\pgfmathsetmacro{\labelDiff}{int((abs(\lenArg)-abs(\lenMult))*100)))}
				\def\placeFromName{left}%
				\def\placeFromMult{right}%
				\def\anchorFromMult{west}%
				\def\anchorFromName{east}%
				\pgfmathparse{int(abs(\yDiff) - abs(\xDiff))}%
				\pgfmathsetmacro{\hvDiff}{\pgfmathresult}
				\ifnum\pgfmathresult<0%
							
				\edef\placeFromName{above}%
				\edef\placeFromMult{below}%
				\ifnum\labelDiff<0%
				\ifnum\intXDiff>0%
				\edef\anchorFromName{north west}%
				\edef\anchorFromMult{south west}%
				\else%
				\edef\anchorFromName{south east}%
				\edef\anchorFromMult{north east}%
				\fi%
				\else%
				\ifnum\intXDiff>0%
				\edef\anchorFromName{north west}%
				\edef\anchorFromMult{south west}%
				\else%
				\edef\anchorFromName{north east}%
				\edef\anchorFromMult{south east}%
				\fi%
				\fi%
				\else%
				\fi%
				\node[\placeFromName, anchor=\anchorFromName] {\fromArg};%
				\node[\placeFromMult, anchor=\anchorFromMult]{\vphantom{\fromArg}\fromMult};%
			},%
			mark=at position \sPos with
			{	
				\node[above, transform shape] {\stereotype};								
			},%
			mark=at position \toPos-0.02 with
			{
				\node (D){};	
				\pgfgetlastxy{\xlast}{\ylast}
				\global\let\oldX\xlast
				\global\let\oldY\ylast				
			},
			mark=at position \toPos with
			{
				\node(E){};
				\pgfmathsetmacro\lenArg{width("\toArg")}				
				\pgfmathsetmacro\lenMult{width("\toMult")}
				\pgfgetlastxy{\xlast}{\ylast}%
				\pgfmathsetmacro{\xDiff}{(abs(\xlast)-abs(\oldX))*100}%
				\pgfmathsetmacro{\yDiff}{(abs(\ylast)-abs(\oldY))*100}%
				\pgfmathsetmacro{\intXDiff}{int((\xlast-\oldX)*100)}
				\pgfmathsetmacro{\labelDiff}{int((\lenArg-\lenMult)*100)))}
				\def\placeToName{left}%
				\def\placeToMult{right}%

				\pgfmathparse{int(abs(\yDiff) - abs(\xDiff))}%
				\ifnum\pgfmathresult<0%
				\edef\placeToName{above}%
				\edef\placeToMult{below}%
				\ifnum\labelDiff<0%
				\ifnum\intXDiff>0%
				\else%
				\fi%
				\else%
				\ifnum\intXDiff>0%
				\else%
				\fi%
				\fi%
				\else%
				\fi%
				\node[\placeToName] {\toArg};				
				\node[\placeToMult]{\vphantom{\toArg}\toMult};
			},%
		},%
	},%
}%
\tikzset{%
	rel label/.style args={#1#2#3}{%
		postaction={ decorate,%
			decoration={ markings,%
				mark=at position #1 with \node[{#2,text height=1.5ex, text depth=0.25ex}] {#3};%
			}%
		}%
	},%
	TLeft/.style={label={[anchor=east, xshift=1ex]left:$\blacktriangleleft$}},%
	TRight/.style={label={[anchor=west, xshift=-1ex]right:\strut$\blacktriangleright$}},%
	TUp/.style={label={[anchor=south, yshift=-1ex]above:$\blacktriangle$}},%
	TDown/.style={label={[anchor=north, yshift=1ex]below:$\blacktriangledown$}},%
	closer pos/.style={},%
	class/.append style={minimum width=6.5em, inner ysep=0.5em, inner xsep=0.1em, minimum height=2.75em, font=\bfseries\footnotesize},%
	assoc label/.style={align=center, font=\scriptsize},%
	aggregation/.style={%
		association={#1},%
		-{Diamond[open, length=1.2em]}%
	},%
	generalization/.style={
		draw,
		-{Triangle[open, length=0.7em, width=0.8em]}%
	},%
	specialization/.style={
		draw,
		-{Triangle[length=0.7em, width=0.8em]}%
	},%
	composition/.style={%
		association={#1},
		-{Diamond[length=1.3em]}
	},
	part composite/.style={%
		association={#1},
		{Straight Barb}-{Diamond[length=1.1em]}
	},
	part reference/.style={%
		association={#1},
		{Straight Barb}-{Diamond[open,length=1.1em]}
	}
}%
\newcounter{samplingidx}%
\newcounter{coordindex}%
\newcounter{constraintidx}%
\tikzset{%
	stepcounter-coord/.code={%
		\stepcounter{coordindex}%
	},%
	stepcounter-sample/.code={%
		\stepcounter{samplingidx}%
	},%
	stepcounter-constraint/.code={%
		\stepcounter{constraintidx}%
	}%
}%
\tikzset{%
	polyline/.style={%
		decoration={%
			markings,%
			mark=between positions 0 and 1 step 0.05 with {%
				\coordinate[stepcounter-coord] (A\thecoordindex) at (0,0);%
			}%
		},%
		postaction=decorate,%
	}%
}%
\tikzset{%
	tangent/.style={%
		decoration={%
			markings,
			mark=%
			at position #1%
			with%
			{%
				\coordinate (tangent point-\pgfkeysvalueof{/pgf/decoration/mark info/sequence number}) at (0pt,0pt);%
				\coordinate (tangent unit vector-\pgfkeysvalueof{/pgf/decoration/mark info/sequence number}) at (1,0pt);%
				\coordinate (tangent orthogonal unit vector-\pgfkeysvalueof{/pgf/decoration/mark info/sequence number}) at (0pt,1);%
			}%
		},%
		postaction=decorate%
	},%
	use tangent/.style={%
		shift=(tangent point-#1),%
		x=(tangent unit vector-#1),%
		y=(tangent orthogonal unit vector-#1)%
	},%
	use tangent/.default=1%
}%
\tikzset{%
	constraint line/.style={%
		decoration={%
			markings,%
			mark=between positions 0 and 0.85 step 0.05 with {%
				\node[stepcounter-constraint, circle, inner sep=1pt, fill=tuRed] (lwr-c-\theconstraintidx) at (0,\lowerconstraints[\theconstraintidx]) {};%
				\draw[tuLightGreen](0,0) -- (lwr-c-\theconstraintidx);%
				\draw[tuLightGreen] (0,0) -- (0,3);%
			}%
		},%
		postaction=decorate,%
	}%
}%
\tikzset{%
	sampling line/.style={%
		decoration={%
			markings,
			mark=between positions 0 and 1 step 0.05 with {%
				\coordinate[stepcounter-sample] (B\thesamplingidx-left) at (0,3);
				\coordinate[] (B\thesamplingidx-right) at (0,-6);
			}			
		},
		postaction=decorate,
	}
}
\tikzset{cross/.style={cross out, draw=black, minimum size=2*(#1-\pgflinewidth), inner sep=0pt, outer sep=0pt},	cross/.default={1pt}}

\pgfdeclarelayer{background}%
\pgfdeclarelayer{pre main}%
\pgfdeclarelayer{foreground}%
\pgfdeclarelayer{lifelines}%
\pgfdeclarelayer{activity}%
\pgfdeclarelayer{connections}%
\pgfsetlayers{background,lifelines,activity, connections,pre main,main,foreground}%

%% file: fusa-sotif-mapping.tikz
\tikzstyle{shadedGreenLight}  = [top color=tuLightGreen20, bottom color=tuLightGreen60,  draw=tuGreen100, thick]%
\tikzstyle{shadedVioletLight}  = [top color=tuLightViolet20, bottom color=tuVioletLight40,  draw=tuVioletLight100, thick]%
\tikzset{%
	diaBlock/.style = {%
		block, shadedGray,%
		inner ysep=0.25em,%
		inner xsep=0.25em,%
		align=center,%
		minimum width=7.5em,%
		minimum height=2.9em,%
		node distance=0.3em and 1em,%
		drop shadow,%
            font=\sffamily\scriptsize%
	},%
	artifact/.style={%
		diaBlock,%
            node distance=0.3em and 0.5em,%
		minimum width=7em,%
		rounded corners=0pt,%
            font=\sffamily\scriptsize%
	}%
}%
\begin{tikzpicture}[every path/.style={%
		thick,
		-{Stealth[round]},
		font=\footnotesize
	}]
	 \node[diaBlock, shadedGreenLight] (first) at (0,0) {Business or \\ Mission Analysis\\Process};
	 \node[diaBlock, shadedOrange, below= of first] (second) {Stakeholder Needs \\ and Requirements\\Definition Process};
	 \node[diaBlock, shadedBlueLight, below= of second] (third) {System \\Requirements\\ Definition Process};
	 \node[diaBlock, shadedVioletLight, below= of third] (fourth) {System\\Architecture\\Definition Process};
	 \node[diaBlock, shadedYellow, below= of fourth] (fifth) {System \\ Design Definition\\Process};
	 \node[artifact, shadedOrange, below right= 0.5em and 3em of first.east] (sixth) {Normative\\Stakeholder\\Requirements};
	 \node[artifact, shadedBlueLight, below= of sixth] (seventh) {Actuator\\Capabilities};
	 \node[artifact, shadedOrange, right= 1em of sixth] (eigth) {Use Cases / \\ Assumptions for \\ Vehicle Behavior};	
	 \node[artifact, shadedBlueLight, below= of eigth] (ninth) {Performance\\Requirements};
	 \node[artifact, shadedVioletLight, below= of ninth] (tenth) {(Preliminary)\\Functional\\Architecture};
	 \node[artifact, shadedYellow, below= of tenth] (11) {(Preliminary)\\Functional\\Design};	
	 \node[artifact, shadedOrange, right= 1em of eigth, minimum width=9em] (12) {Operational\\Design \\ Domain (ODD)};
	 \node[artifact, shadedOrange, below=of 12, minimum width=9em] (13) {Interfaces to \&\\Interaction with \\ External Entities};
	 \node[artifact, shadedVioletLight, below= of 13, xshift=1em] (14) {Physical \\Architecture};
	 \node[artifact, shadedYellow, below= of 14] (15) {Physical\\Design};
	
	\begin{pgfonlayer}{pre main}
		\node[fit=(first)(fifth), draw=tuGrey, fill=lightgray!30, inner xsep=0.7em, inner ysep=0.9em, yshift=0.33em, thick, rounded corners=2pt, label={[anchor=north, font=\sffamily\footnotesize]:\textbf{ISO 15288}}] {};
		\node [draw, fill=tuLightBlue20!50,fill opacity=90, fit=(eigth)(15), inner sep=1em, inner ysep=1.25em, yshift=0.4em, rounded corners=2pt, label={[anchor=north east, font=\bfseries\sffamily\footnotesize]north east:Specification\,\&\,Design}] (fuspec) {};	
		\node [draw, fit=(sixth)(11), fill=tuLightViolet20!30, fill opacity=90, inner sep=0.5em, rounded corners=2pt, inner xsep=1.5em, xshift=0.5em, label={[anchor=south west,align=center, font=\bfseries\sffamily\footnotesize]south west:Item\\Definition\\(Vehicle Level)}] (itemdef) {};
	\end{pgfonlayer}
	
	\begin{pgfonlayer}{background}
		\node[fit=(itemdef)(fuspec), draw=tuGrey, fill=lightgray!30, rounded corners=2pt, inner sep=0.7em, inner ysep=1.4em, yshift=0.25em, label={[anchor=north, font=\bfseries\sffamily\footnotesize]:ISO~26262 / ISO~21448}] (safety) {};
	\end{pgfonlayer}%
	\draw[arrow](first) to (\tikztostart -| safety.west);%
	\draw[arrow](second) to (\tikztostart -| safety.west);%
	\draw[arrow](third) to (\tikztostart -| safety.west);%
	\draw[arrow](fourth) to (\tikztostart -| safety.west);%
	\draw[arrow](fifth) to (\tikztostart -| safety.west);%
\end{tikzpicture}%